\documentclass[preprintnumbers,nofootinbib,superscriptaddress,aps,prd]{revtex4}

\usepackage{amsmath,amssymb,bm}
\usepackage{psfrag}
\usepackage{graphicx}

\newcommand{\scetopi}{\text{1PI}_{\text{SCET}}}
\newcommand{\scetpim}{\text{PIM}_{\text{SCET}}}

\begin{document}

\preprint{MZ-TH/11-12}
\preprint{ZU-TH 11/11}

\title{The top-pair forward-backward asymmetry beyond NLO}

\author{Valentin Ahrens}
\affiliation{Institut f\"ur Physik (THEP), Johannes Gutenberg-Universit\"at, D-55099
  Mainz, Germany}
\author{Andrea Ferroglia}
\affiliation{New York City College of Technology, 300 Jay Street, Brooklyn, NY 11201, USA}
\author{Matthias Neubert}
\author{Ben D.~Pecjak}
\affiliation{Institut f\"ur Physik (THEP), Johannes Gutenberg-Universit\"at, D-55099
  Mainz, Germany}
\author{Li Lin Yang}
\affiliation{Institute for Theoretical Physics, University of Z\"urich, CH-8057 Z\"urich,
  Switzerland}

\date{\today}

\begin{abstract}
  \noindent
  We make use of recent results in effective theory and higher-order
  perturbative calculations to improve the theoretical predictions of the QCD
  contribution to the top-quark pair production forward-backward asymmetry at
  the Tevatron. In particular, we supplement the fixed-order NLO calculation
  with higher-order corrections from soft gluon resummation at NNLL accuracy
  performed in two different kinematic schemes, which allows us to make
  improved predictions for the asymmetry in the $p\bar p$ and $t\bar t$ rest
  frames as a function of the rapidity and invariant mass of the $t\bar t$
  pair. Furthermore, we provide binned results which can be compared with the
  recent measurements of the forward-backward asymmetry in events with a large
  pair invariant mass or rapidity difference.   Finally, we calculate
  at NLO+NNLL order the top-quark charge asymmetry at the LHC as a function of
  a lower rapidity cut-off for the top and antitop quarks.
\end{abstract}

\maketitle

\section{Introduction} 

The forward-backward (FB) asymmetry in top-quark pair production in
proton-antiproton collisions is an observable which originates from the
difference in the production rates for top quarks in the forward and backward
hemispheres \cite{Kuhn:1998jr, Kuhn:1998kw}.
The total FB asymmetry was measured by the CDF and D0 collaborations at the
Tevatron \cite{:2007qb, Aaltonen:2008hc, Aaltonen:2011kc}. The measurement can
be carried out in the laboratory frame ($p\bar{p}$ frame) as well as in the
center-of-mass frame of the top-quark pair ($t\bar{t}$ frame). The asymmetries
in the two frames are defined as
\begin{align}
  A^{i}_{\text{FB}} &= \frac{N(y^i_t>0)-N(y^i_t<0)}{N(y^i_t>0)+N(y^i_t<0)} \, ,
\end{align}
where $N$ is the number of events, $i = p\bar{p} \, (t\bar{t})$ indicates the
laboratory frame ($t\bar{t}$ frame), and $y^i_t$ is the top-quark rapidity in frame $i$.
The measurements obtained by the CDF collaboration using 5.3~fb$^{-1}$ of
data are \cite{Aaltonen:2011kc} 
\begin{align}
  \label{eq:exp}
  A^{p\bar{p}}_{\text{FB}} &= (15.0 \pm 5.5)\% \quad (\text{$p\bar{p}$ frame}) \, ,
  \nonumber
  \\
  A^{t\bar{t}}_{\text{FB}} &= (15.8 \pm 7.5)\% \quad (\text{$t\bar{t}$ frame}) \, .
\end{align}
The quoted uncertainties are derived from a combination of statistical and systematic errors.

The production of top-quark pairs at hadron colliders is dominated by QCD. At
the Tevatron,  the charge conjugation invariance of the strong interaction
implies that the difference in the production of top quarks in the
forward and backward hemispheres is equivalent to the difference in the production of top and
antitop quarks in the forward hemisphere. Therefore, in QCD the FB asymmetry is equivalent
to the charge asymmetry. QCD predicts a non-vanishing contribution to the FB asymmetry
starting at order $\alpha_s^3$ in the squared amplitude. A contribution to the asymmetry
arises if, in the interference of one-loop and tree-level diagrams, the top-quark
fermionic line and the light-quark fermionic line are connected by three gluons
\cite{Kuhn:1998kw}. The same is true in the case of the interference of two tree-level
diagrams with three particles in the final state. The dominant effect is from
the quark-antiquark annihilation channel, while a further, numerically small
contribution to the asymmetry at order $\alpha_s^3$ originates from the 
flavor excitation channel $gq\to t \bar{t} X$, 
where $X$ indicates additional partons in the final state. The gluon 
fusion channel does not contribute to the FB asymmetry at any 
order in perturbation theory, due to the fact that the gluon distribution is the same for 
protons and antiprotons.

The total FB asymmetry predicted by QCD at the first non-vanishing order (which, for reasons
discussed later in this paper, we will indicate as next-to-leading order (NLO)) is lower than the one measured at
the Tevatron. Two recent evaluations using the formulas in \cite{Kuhn:1998kw} report the
following values \cite{Ahrens:2010zv,Ahrens:2011mw}, obtained using MSTW2008 NLO parton distribution functions (PDFs)
\cite{Martin:2009iq},
\begin{align}
  \label{eq:NLO}
  A^{\text{$p\bar{p}$, NLO}}_{\text{FB}} &= (4.8^{+0.5}_{-0.4})\% \quad (\text{$p\bar{p}$
    frame}) \, , \nonumber
  \\
  A^{\text{$t\bar{t}$, NLO}}_{\text{FB}} &= (7.4^{+0.7}_{-0.6})\% \quad (\text{$t\bar{t}$
    frame}) \, .
\end{align}
The central values quoted in (\ref{eq:NLO}) refer to the choice $\mu_f =m_t$, where
$\mu_f$ is the factorization scale and $m_t$ the top-quark mass, which is set to $m_t=173.1$~GeV. 
The errors originate from the scale variation in the
range $m_t/2 \le \mu_f \le 2 m_t$. These choices will also be adopted in the rest of the
paper. Electroweak corrections enhance the prediction for the asymmetry by less than 
10\% of the central value \cite{Kuhn:1998kw, Bernreuther:2010ny},  and to separate 
uncertainties coming from electroweak 
calculations we do not include these corrections in our numerical results.
As one can see, the discrepancy between the
theory prediction and the experimental measurement in the $p\bar{p}$ frame is less than two
standard deviations (2$\sigma$), while in the $t\bar{t}$ frame the two values agree within
$ \sim 1 \sigma$, although the central value of the experimental measurement is a bit higher.

In \cite{Aaltonen:2011kc}, the CDF collaboration measured the FB asymmetry in the
$t\bar{t}$ frame as a function of the top-pair invariant mass $M_{t\bar{t}}$. After
grouping the events in two bins corresponding to $M_{t\bar{t}} \le 450$~GeV and $M_{t\bar{t}} \ge 450$~GeV,
they found the asymmetry in the latter bin to be
\begin{align}
  A^{t\bar{t}}_{\text{FB}} \left(M_{t\bar{t}} \ge 450 \, \mathrm{GeV} \right) 
  = (47.5 \pm 11.4)\% \, ,
\end{align}
which is more than 3$\sigma$ higher than the stated  theoretical NLO prediction in \cite{Aaltonen:2011kc} of $(8.8\pm1.3)\%$
obtained using the MCFM program~\cite{Campbell:1999ah}.  A measurement of the FB asymmetry 
in two bins of the rapidity difference $y_t-y_{\bar t}$ was also performed and again
in that case the higher bin shows a tension with the NLO QCD prediction, 
although  with larger experimental errors.   
Many attempts to explain these results in terms of new physics scenarios have
been made, see e.g.\ 
\cite{Cheung:2011qa, Bai:2011ed, Berger:2011ua, Bhattacherjee:2011nr, Barger:2011ih,
  Patel:2011eh, Ligeti:2011vt, Grinstein:2011yv, 
Gresham:2011pa, Jung:2011zv, Shu:2011au}. The task is
complicated by the fact that the new physics contributions should not spoil the good
agreement between theory and measurements for the total pair-production cross
section and the differential distribution in the pair invariant mass.

Recently, we have calculated the top-pair invariant-mass distribution and the top-quark
rapidity and transverse-momentum distributions in renormalization-group (RG) improved
perturbation theory, which incorporates the resummation of logarithmic
corrections that become numerically large in the limit of soft-gluon emission
\cite{Ahrens:2010zv, Ahrens:2011mw}. These calculations were carried out at
next-to-next-to-leading-logarithmic (NNLL) accuracy. 
The resummed distributions were then
matched to the NLO results to obtain predictions which have
NLO+NNLL accuracy.
Alternatively, the resummed results at NNLL order can be employed to generate approximate next-to-next-to leading (NNLO)
predictions \cite{Ahrens:2010zv, Ahrens:2011mw}. 
We note that working at approximate NNLO accuracy
does not alter in a fundamental way  any of the results  obtained  within the NLO+NNLL framework. On the other hand, we believe that
in certain kinematic regions, for instance at large pair invariant mass,
resummation is important. For these reasons, and for the
sake of simplicity, in the following we focus  on NLO+NNLL calculations. 
By integrating the differential distributions it is then straightforward
to calculate the top-quark FB asymmetry both in the laboratory frame and in the $t\bar{t}$
frame. Starting from the double differential cross section in the pair invariant mass
and scattering angle, we can also compute the
$M_{t\bar{t}}$ and rapidity-dependent asymmetries, which can be compared to experimental measurements.
The main goal of this letter is to present the result of these calculations of the total and differential
FB asymmetries in a systematic way.

At the Large Hadron Collider (LHC), the symmetry of the
$pp$ initial state dictates that the rapidity distributions of the top and antitop
quarks are symmetric and that the FB asymmetry vanishes.  However, it was 
observed in \cite{Kuhn:1998kw} that at the LHC top quarks are 
preferably produced at larger rapidities than  antitop quarks in the
laboratory frame.
Like the FB asymmetry at the Tevatron, this rapidity-dependent 
charge asymmetry is generated at order $\alpha_s^3$ in the squared amplitude,
mainly through the asymmetric part of the quark-antiquark annihilation
channel.  Therefore, potential new physics contributions would effect
these two quantities in a correlated way, and the higher collider
energy at the LHC gives it better access to distortions at 
higher rapidities.   
As a final application of our formalism, we  evaluate at NLO+NNLL
order the partially integrated charge asymmetry at the LHC, giving 
results as a function of a lower cut-off on the top and antitop
rapidities in the laboratory frame.

\section{FB Asymmetry in the Laboratory Frame}
\label{sec:ppasy}

\subsection{Total asymmetry}

The FB asymmetry in the laboratory frame can be calculated starting from the top-pair
production cross section differential with respect to the top-quark transverse momentum
$p_T$ and rapidity $y_t$. To do so,
it is convenient to first define a total and differential asymmetric cross section via
\begin{align}
  \label{eq:fbasympp}
  \Delta\sigma^{p\bar{p}}_{\text{FB}} &\equiv \int_0^{y^+_t} dy_t \left[
    \int_0^{p_T^{\text{max}}} dp_T \, \frac{d^2 \sigma^{p\bar{p} \to t
        X_{\bar{t}}}}{dp_Tdy_t} - \int_0^{p_T^{\text{max}}} dp_T \,\frac{d^2
      \sigma^{p\bar{p} \to t X_{\bar{t}}}}{dp_Td\bar{y}_t} \Bigg|_{\bar{y}_t=-y_t} \right]
  \nonumber
  \\
  &\equiv \int_0^{y^+_t} dy_t \left[ \left( \frac{d\sigma}{dy_t} \right)_F - \left(
      \frac{d\sigma}{dy_t} \right)_B \right]\equiv \int_0^{y^+_t} dy_t  \frac{d \Delta\sigma^{p\bar{p}}_{\text{FB}}}{dy_t} \,.
\end{align}
Here
\begin{align}
\label{eq:yplimits}
  y^+_t = \frac{1}{2} \ln\frac{1+\sqrt{1-4m_t^2/s}}{1-\sqrt{1-4m_t^2/s}} \quad \text{and}
  \quad p_T^{\text{max}} = \frac{\sqrt{s}}{2}\sqrt{ \frac{1}{\cosh^2 y_t}-\frac{4m_t^2}{s}}
  \, ,
\end{align}
where $s$ is the square of the hadronic center-of-mass energy. To obtain the FB asymmetry
in the laboratory frame one needs to calculate the ratio of the asymmetric cross section
in (\ref{eq:fbasympp}) to the total cross section:
\begin{align}
  \label{eq:fbaA}
  A^{p\bar{p}}_{\text{FB}} = \frac{\Delta\sigma^{p \bar{p}}_{\text{FB}}}{\sigma} \,.
\end{align}

In our phenomenological analysis we will consider two levels of perturbative 
precision for the asymmetric cross section and FB asymmetry.  The first 
involves the differential cross section at NLO in fixed-order perturbation
theory, the second the NLO calculation supplemented with soft-gluon
resummation to NNLL order.  In the 
laboratory frame, the resummed calculations are carried out using the  $\scetopi$ scheme 
introduced in \cite{Ahrens:2011mw}.  Such predictions resum a class of
logarithms which become large in the limit where   $s_4 =
(p_{\bar{t}}+k)^2-m_t^2$ is small ($p_{\bar{t}}$ indicates the momentum of the antitop
quark, and $k$ the total momentum of additional partons in the final state),
in which case real gluon emission is soft although the top pair is not 
produced at rest. While the phase-space integrals in (\ref{eq:fbasympp})
are in general sensitive to regions where $s_4$ is not small, the threshold
region is dynamically enhanced due to the sharp fall-off of the PDFs away
from small values of $s_4$.  In fact, it was shown in  \cite{Ahrens:2011mw}
that the leading terms in the $\scetopi$ threshold limit reproduce essentially the entire 
NLO correction to the differential cross section in the quark channel,
which implies that resumming the corrections which become large in this limit 
is an improvement on the fixed-order calculation.  While in the fixed-order 
counting the asymmetric cross section first arises from the NLO calculation
of the differential cross section, in the resummed counting it first 
appears at NLL order. The NLO+NNLL calculation is thus a refinement on the 
leading term, and will be considered our best prediction.  

Before illustrating our results for the FB asymmetry, we need to clarify an
important point concerning our convention for counting orders in the
perturbative expansion.  We calculate $A_{\text{FB}}$ itself as a perturbative
expansion in $\alpha_s$, using a fixed-order or logarithmic counting as
appropriate. For example, the first non-vanishing contribution to the
asymmetry in fixed-order perturbation theory is obtained by calculating the
numerator in (\ref{eq:fbaA}) to order $\alpha_s^3$ and the denominator to
order $\alpha_s^2$. The resulting asymmetry is of order $\alpha_s$, which we
will refer to as NLO, with reference to the order at which the differential
distributions in (\ref{eq:fbasympp}) are calculated relative to
$\alpha_s^2$. Similarly, in RG-improved perturbation theory, the first
non-vanishing contribution to $A_{\text{FB}}$ is obtained by calculating both
the numerator and the denominator at next-to-leading-logarithmic (NLL) order;
the resulting asymmetry is then counted as NLL. This convention was already
adopted in \cite{Ahrens:2010zv, Ahrens:2011mw}. There are two counting schemes
in the literature (in fixed-order perturbation theory) which are different
from ours. The first one also treats $A_{\text{FB}}$ itself as a perturbative
expansion, but counts the order $\alpha_s$ contribution as LO, and so on, as
in \cite{Kuhn:1998jr, Kuhn:1998kw}. The second one treats the numerator and
the denominator as separate perturbative series and does not further expand
the ratio, as adopted in the quoted MCFM results in \cite{Aaltonen:2011kc}.
We note that while the first scheme differs from ours only by name, the second
scheme leads to different numerical results.  As discussed in
\cite{Ahrens:2010zv}, the NLO+NNLL results are considerably more stable with
respect to the choice of scheme than the NLO results.

\begin{table}[h]
  \centering
  \begin{tabular}{|l|c|c|c|c|c|c|}
    \hline
    & \multicolumn{2}{c|}{MSTW2008} & \multicolumn{2}{c|}{CTEQ6.6} & \multicolumn{2}{c|}{NNPDF2.1}
    \\ \cline{2-7}
    & $\Delta\sigma^{p\bar p}_{\rm FB}$ [pb] &  $A^{p\bar p}_{\text{FB}}$ [\%]
    & $\Delta\sigma^{p\bar p}_{\rm FB}$ [pb] &  $A^{p\bar p}_{\text{FB}}$ [\%]
    & $\Delta\sigma^{p\bar p}_{\rm FB}$ [pb] &  $A^{p\bar p}_{\text{FB}}$ [\%]
    \\ \hline
    NLO 
    & 0.260{\footnotesize $^{+0.141+0.020}_{-0.084-0.014}$}
    & 4.81{\footnotesize $^{+0.45+0.13}_{-0.39-0.13}$}
    & 0.256{\footnotesize $^{+0.135}_{-0.082}$}
    & 4.69{\footnotesize $^{+0.44}_{-0.38}$}
    & 0.269{\footnotesize $^{+0.144}_{-0.086}$}
    & 4.82{\footnotesize $^{+0.47}_{-0.38}$}
    \\ \hline 
    NLO+NNLL
    & 0.312{\footnotesize $^{+0.027+0.023}_{-0.035-0.019}$}
    & 4.88{\footnotesize $^{+0.20+0.17}_{-0.23-0.18}$}
    & 0.319{\footnotesize $^{+0.026}_{-0.037}$}
    & 4.79{\footnotesize $^{+0.17}_{-0.25}$}
    & 0.335{\footnotesize $^{+0.029}_{-0.039}$}
    & 4.93{\footnotesize $^{+0.22}_{-0.24}$}
    \\ \hline
\end{tabular}
\caption{\label{tab:ppbar} 
  The asymmetric cross section and FB asymmetry in the $p\bar{p}$ frame. The first error refers 
to perturbative uncertainties estimated through scale variations as explained in the text,
and the second error in the MSTW2008 case is the PDF uncertainty.}
\end{table}

Our results for the total FB asymmetry in the lab frame
are shown in Table~\ref{tab:ppbar}.  As
explained above, to calculate each entry in the table the numerator and the
denominator in (\ref{eq:fbaA}) are evaluated at the order indicated in the
leftmost column, and then the ratio itself is expanded in powers of $\alpha_s$
up to the appropriate order.  The central values  are
obtained by fixing the factorization scale at $\mu_f =m_t$, and the scale
uncertainties are estimated by varying $\mu_f$ between $m_t/2$ and
$2m_t$.\footnote{Although we use $m_t=173.1$~GeV throughout the analysis, the
asymmetry is rather stable under the exact choice of $m_t$.  For instance, at
$m_t=160$~GeV the default value for the asymmetric cross section at NLO with
MSTW2008 PDFs changes to $\Delta\sigma_{\rm FB}^{p\bar p}=0.384$~pb, but the
asymmetry itself changes only by a small amount to $A^{p\bar
p}_{\text{FB}}=4.67\%$.}  In the resummed calculations also
the hard and soft scales are varied according to the procedure described in
\cite{Ahrens:2011mw}, and the effect of these variations is included in the
scale uncertainty shown in the tables.  In the first column of the table
we use MSTW2008 PDFs and
estimate the PDF uncertainties by iterating through the 90\% confidence level (CL) sets.  Here
and throughout this work we use NLO PDFs for the NLO prediction, and switch to
NNLO PDFs for the NLO+NNLL calculation.  The PDF uncertainties for the
asymmetry, expressed as a percentage of the central values, are about half as
large as those for the asymmetric cross section. This is due to cancellations
in the ratio.  In addition to results with MSTW2008 PDFs, we also show those
obtained using CTEQ6.6 \cite{Nadolsky:2008zw} and NNPDF2.1 \cite{Ball:2011mu}
PDFs.  In those cases the PDFs are based on a NLO fit so that the same set is
used in both the NLO and the NLO+NNLL calculations.  We note that the results
for the asymmetry obtained with the different PDF sets are well within the PDF
uncertainties estimated through the MSTW2008 results.

Adding soft-gluon resummation at NNLL accuracy produces results for the
asymmetric cross section which are numerically consistent with the NLO results
for $\mu_f = m_t$, while the scale uncertainty is reduced by more than a factor of 2. The
central value for the FB asymmetry does not change significantly with respect
to the NLO predictions, and also in this case the scale uncertainties are
reduced.  We can therefore conclude that the discrepancy between theory and
experiment cannot be explained with the effect of higher-order QCD corrections
on the theory side, at least not those related to soft-gluon resummation.

The NLO+NNLL calculation represents the most accurate determination of
the QCD contribution to the asymmetry that can be obtained at present.
However, it is important to keep in mind the uncertainties related to
yet higher-order corrections and how they could be reduced.
Concerning soft gluon resummation, the most important improvement beyond
NNLL accuracy would be the calculation of the hard and soft matching
functions at NNLO order.  This requires the soft plus virtual
corrections at NNLO in QCD, and would fix the $\delta$-function
contribution to the differential cross section at that order.  In that
case, the further improvement of matching the resummed results to
fixed order would require power-suppressed effects related to hard
gluon emission.  The arguments based on the dynamical enhancement of
the threshold region and the confirmation of this mechanism through
the numerical results at NLO, imply that power corrections to the soft
limit are small, so we expect the more important effect to be the
calculation of the soft plus virtual corrections.  We estimate
uncertainties related to both types of corrections through the
standard method of scale variations and the numerical results indicate
that these higher-order effects are moderate. However, this statement
can of course never be certain without the actual calculation of the
higher-order pieces.  Similar comments apply to all other quantities
obtained in this work.

\subsection{Rapidity-dependent asymmetry}

\begin{figure}
\begin{center}
\begin{tabular}{ll}
\psfrag{y}[][][1][90]{$d\Delta\sigma^{p\bar{p}}_{\text{FB}}/dy_t$\;[pb]}
\psfrag{x}[]{$y_t$}
\psfrag{d}[][][0.90]{$\sqrt{s}=1.96$\,TeV}
\includegraphics[width=0.5\textwidth]{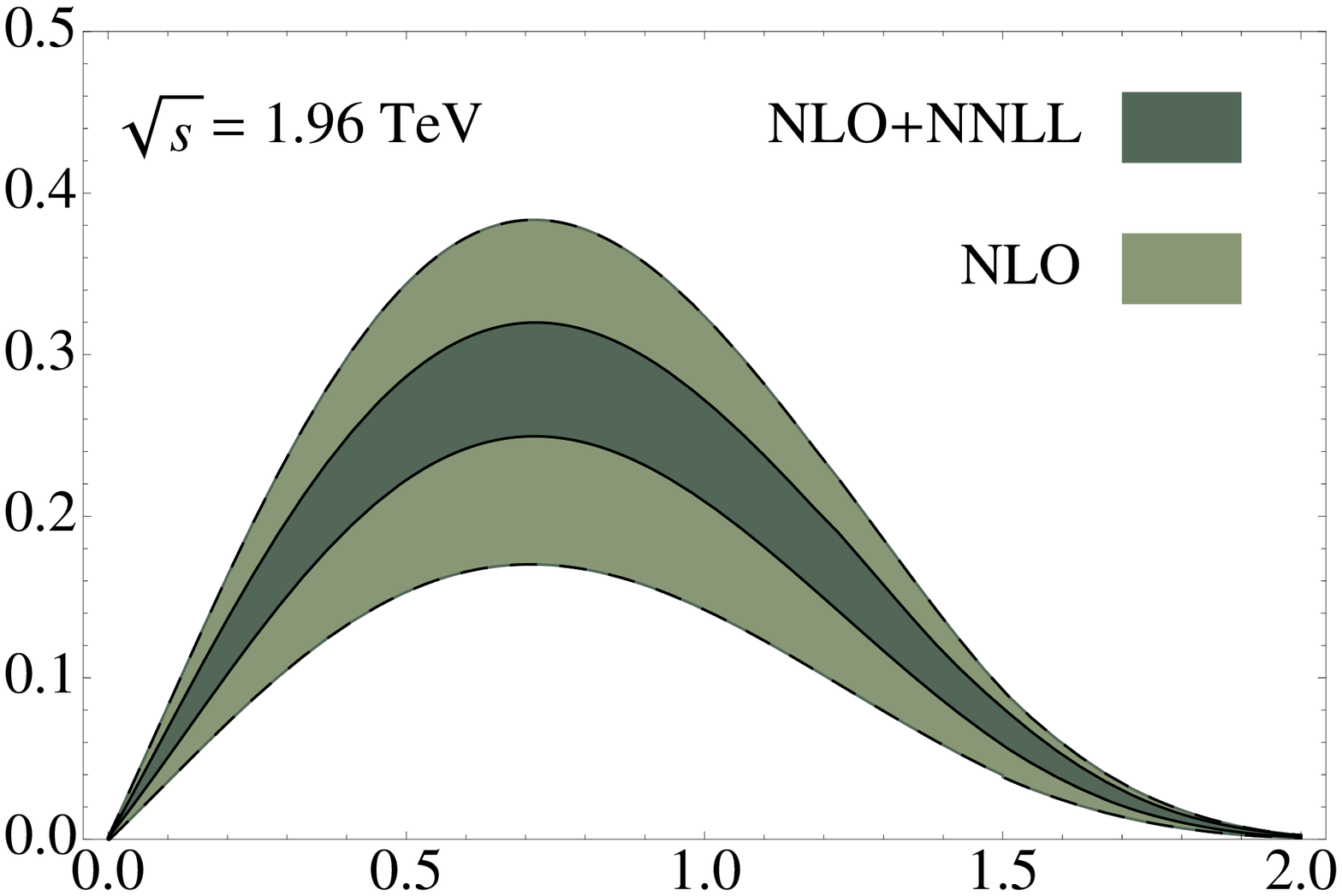}
\psfrag{y}[][][1][90]{$A^{p \bar{p}}_{\text{FB}}(y_t)$}
\psfrag{x}[]{$y_t$}
\psfrag{d}[][][0.90]{$\sqrt{s}=1.96$\,TeV}
\includegraphics[width=0.5\textwidth]{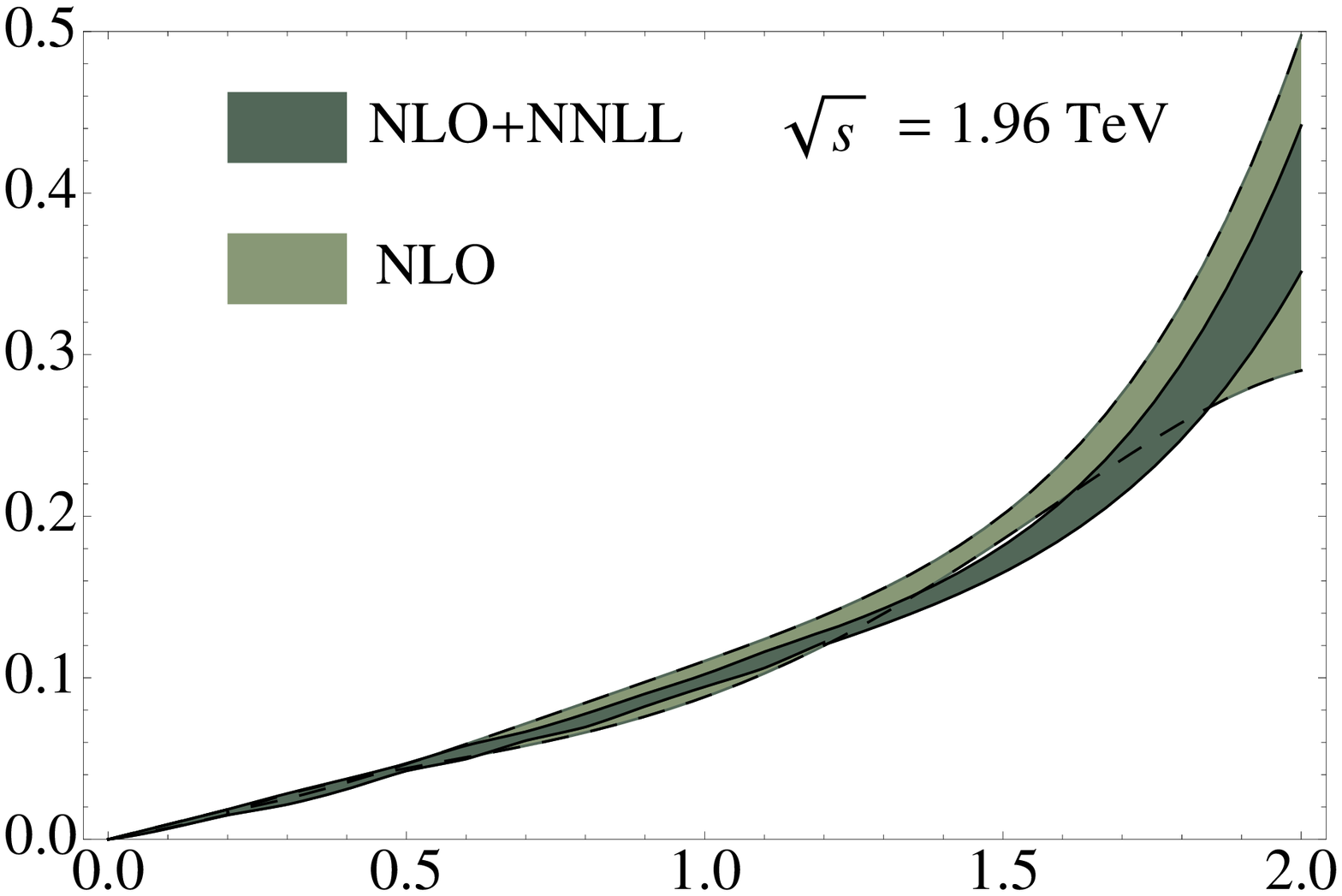}
\end{tabular}
\end{center}
\vspace{-2mm}
\caption{\label{fig:rapidity} Left: The asymmetric differential cross section $ d\Delta\sigma^{p
      \bar{p}}_{\text{FB}}/dy_t$. Right: The asymmetry $A^{p \bar{p}}_{\text{FB}}(y_t)$.
  The bands show the uncertainties related to scale variation as explained in the text.}
\end{figure}

As experimental measurements become more precise, differential quantities such as the
rapidity-dependent asymmetry can be compared with theoretical predictions. Using
quantities defined in (\ref{eq:fbasympp}), we can write this differential asymmetry as
\begin{align}
  \label{eq:fbaYdiff}
  A^{p\bar p}_{\text{FB}}(y_t) 
  = \frac{\displaystyle
    \left(\frac{d\sigma}{dy_t}\right)_F - \left(\frac{d\sigma}{dy_t}\right)_B}{\displaystyle
    \left(\frac{d\sigma}{dy_t}\right)_F + \left(\frac{d\sigma}{dy_t}\right)_B} =
    \frac{\displaystyle \frac{d \Delta \sigma^{p\bar{p}}_{\text{FB}}}{dy_t}}{ \displaystyle \left(\frac{d\sigma}{dy_t}\right)_F + \left(\frac{d\sigma}{dy_t}\right)_B }\, .
\end{align}
In Figure~\ref{fig:rapidity} we show results for the differential asymmetric cross section
and the FB asymmetry as functions of the rapidity at NLO and NLO+NNLL order, using
MSTW2008 PDFs. Here and below, the differential cross sections at NLO are obtained using a
private NLO version of MadGraph \cite{Frederix:2009yq}. The bands refer to uncertainties
associated with scale variations. One observes that the NLO+NNLL band for the asymmetric
differential cross section displayed in the left-hand panel is contained within the NLO band over the entire range of $y_t$
values shown in the figure, and that the scale uncertainty of the NLO+NNLL asymmetric
cross section is smaller than the scale uncertainty obtained in the NLO
calculation. On the other hand, the differences between the error bands in 
the NLO and NLO+NNLL  results for the FB asymmetry shown in the right-hand
panel  are much smaller due to cancellations  in the ratio which make the NLO FB asymmetry
considerably more stable than the corresponding asymmetric cross section. 
We will encounter this feature repeatedly in the
$t\bar t$-frame calculations which follow. One observes that the form of the
rapidity-dependent asymmetry, which is an increasing function with respect to $y_t$, is
very stable under higher-order corrections.

\section{FB Asymmetry in the $\bm{t \bar{t}}$ Frame}

\subsection{Total asymmetry}

Our studies of the FB asymmetry in the $t\bar t$ rest frame will use as a
fundamental quantity the top-pair 
production cross section differential with respect to the pair invariant
mass and the top-quark scattering angle in that frame.   
We thus define  an asymmetric cross section as
\begin{align}
  \label{eq:fbasymtt}
  \Delta\sigma^{t\bar{t}}_{\text{FB}} &\equiv \int_{2m_t}^{\sqrt{s}} dM_{t\bar{t}} \left[
    \int_0^1 d\cos\theta \, \frac{d^2\sigma^{p\bar{p} \to t\bar{t}
        X}}{dM_{t\bar{t}}d\cos\theta} - \int_{-1}^0 d\cos\theta \,
    \frac{d^2\sigma^{p\bar{p} \to t\bar{t} X}}{dM_{t\bar{t}}d\cos\theta} \right] \nonumber
  \\
  &\equiv \int_{2m_t}^{\sqrt{s}} dM_{t\bar{t}} \left[ \left( \frac{d\sigma}{dM_{t\bar{t}}}
    \right)_F - \left( \frac{d\sigma}{dM_{t\bar{t}}} \right)_B \right]
    \equiv \int_{2m_t}^{\sqrt{s}} dM_{t\bar{t}} \frac{d \Delta \sigma^{t\bar{t}}_{\text{FB}}}{dM_{t\bar{t}}} ,
\end{align}
and the total FB asymmetry  in the $t\bar{t}$ frame is then given by
\begin{align}
  A^{t\bar{t}}_{\text{FB}} = \frac{\Delta\sigma^{t\bar{t}}_{\text{FB}}}{\sigma} \,.
\end{align}

As in the previous section, we will study the FB asymmetry at both NLO and NLO+NNLL
accuracy. In the $t\bar t$ frame, the resummed calculations are carried out using the
$\scetpim$ scheme introduced in \cite{Ahrens:2010zv}. The differential cross section
calculated in that work includes the resummation of partonic threshold logarithms up to
NNLL order directly in momentum space, and thus extends the previous results from
\cite{Almeida:2008ug}, which included resummation to NLO+NLL accuracy in moment space.
Such predictions resum a tower of logarithms which become large when the square of the
pair invariant mass is close to the partonic center-of-mass energy: $z =
M^2_{t\bar{t}}/\hat{s}\to 1$. In this limit real gluon emission is soft although, as in
the case of 1PI kinematics, the top quarks are not necessarily produced at rest. While the
soft limit $z\to 1$ can be enforced kinematically by the restriction to large pair
invariant mass, the phase-space integrals in (\ref{eq:fbasymtt}) are in general sensitive
to regions where this is not the case. However, the analysis of \cite{Ahrens:2010zv}
showed that events near the partonic threshold provide the numerically dominant
contributions to the differential cross section, even at low values of the invariant mass
near the peak of the distribution, due to dynamical threshold enhancement from the PDFs
\cite{Becher:2007ty}, which indicates that resumming threshold logarithms is an
improvement on the fixed-order calculation. The suppression of power corrections to the
soft limit is also backed up by the negligible numerical difference between the
calculation of the asymmetry in the $t\bar t$ frame and the partonic center-of-mass frame.
The results in these two frames coincide in the threshold limit $z\to 1$, and differ by
only about 1\% at NLO \cite{Ahrens:2010zv}, due to very small corrections from hard gluon
emission.\footnote{In \cite{Ahrens:2010zv}, results at NLO+NNLL order for the total
  asymmetry were actually given in the partonic center-of-mass frame. In the present work,
  we have eliminated this small mismatch with the experimental measurements by working in
  the $t\bar t$ frame.}
  
\begin{table}[h]
  \centering
  \begin{tabular}{|l|c|c|c|c|c|c|}
    \hline
    & \multicolumn{2}{c|}{MSTW2008} & \multicolumn{2}{c|}{CTEQ6.6} & \multicolumn{2}{c|}{NNPDF2.1}
    \\ \cline{2-7}
    & $\Delta\sigma^{t\bar t}_{\rm FB}$ [pb] & $A^{t\bar t}_{\text{FB}}$ [\%]
    & $\Delta\sigma^{t\bar t}_{\rm FB}$ [pb] & $A^{t\bar t}_{\text{FB}}$ [\%]
    & $\Delta\sigma^{t\bar t}_{\rm FB}$ [pb] & $A^{t\bar t}_{\text{FB}}$ [\%]
    \\ \hline
    NLO 
    & 0.395{\footnotesize $^{+0.213+0.028}_{-0.128-0.021}$} 
    & 7.32{\footnotesize $^{+0.69+0.18}_{-0.59-0.19}$}
    & 0.389{\footnotesize $^{+0.205}_{-0.123}$} 
    & 7.14{\footnotesize $^{+0.67}_{-0.54}$}
    & 0.411{\footnotesize $^{+0.218}_{-0.131}$} 
    & 7.36{\footnotesize $^{+0.70}_{-0.58}$}
    \\ \hline 
    NLO+NNLL
    & 0.448{\footnotesize $^{+0.080+0.030}_{-0.071-0.026}$} 
    & 7.24{\footnotesize $^{+1.04+0.20}_{-0.67-0.27}$}
    & 0.461{\footnotesize $^{+0.083}_{-0.073}$} 
    & 7.16{\footnotesize $^{+1.05}_{-0.68}$}
    & 0.486{\footnotesize $^{+0.088}_{-0.078}$} 
    & 7.39{\footnotesize $^{+1.08}_{-0.69}$}
    \\ \hline  
  \end{tabular}
  \caption{\label{tab:ttbar} The asymmetric cross section and FB asymmetry in the $t\bar{t}$ rest frame. The first error refers to perturbative uncertainties estimated through scale
    variations, and the second error in the MSTW2008 case is the PDF
    uncertainty.}
\end{table}

Our numerical results for the total asymmetric cross section and FB asymmetry 
are summarized in Table~\ref{tab:ttbar}.  As was 
the case in the laboratory frame, the scale uncertainties in the asymmetric cross section
are roughly halved at NLO+NNLL order compared to NLO.  The scale uncertainties in
the FB asymmetry, on the other hand, actually increase slightly after
adding the resummation, while the central values are nearly unchanged.  We note however
that the resummed results are more stable with respect to the scheme for expanding the ratio
defining the FB asymmetry \cite{Ahrens:2010zv}, and in that sense are more reliable than the NLO 
predictions.  Moreover, the resummed results for both the asymmetric and total cross sections
are more stable under scale variations than their fixed-order counterparts.  One should therefore
be cautious of the rather small scale uncertainties in the NLO calculation of the FB asymmetry,
which result from large cancellations in the ratio not observed in the resummed result.    
We again show the PDF uncertainties using the MSTW2008 PDFs at
90\% CL, and the central values and scale uncertainties from the CTEQ6.6 and
NNPDF2.1 sets.  All comments from the previous section 
concerning the reduction of PDF errors in the FB asymmetry compared to the 
asymmetric cross section, and the good agreement between the different PDF
sets, are also true in this case.  

\subsection{Invariant-mass dependent asymmetry}

\begin{figure}
\begin{center}
\begin{tabular}{ll}
\psfrag{y}[][][1][90]{$d\Delta\sigma^{t\bar{t}}_{\text{FB}}/dM_{t\bar{t}}$\;[fb/GeV]}
\psfrag{x}[]{$M_{t\bar{t}}$[GeV]}
\psfrag{d}[][][0.90]{$\sqrt{s}=1.96$\,TeV}
\includegraphics[width=0.5\textwidth]{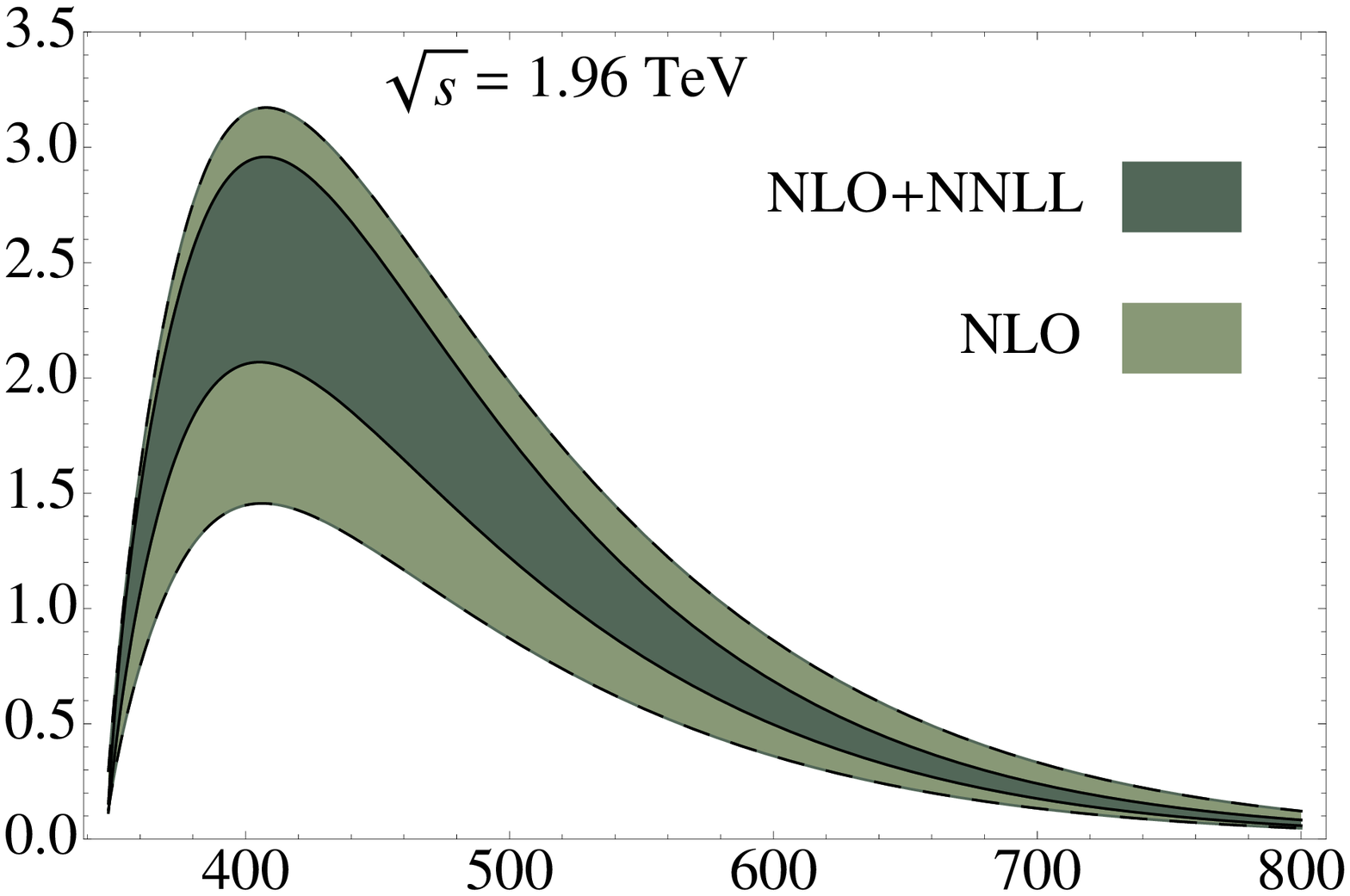}
\psfrag{y}[][][1][90]{$A^{t \bar{t}}_{\text{FB}}(M_{t\bar t})$}
\psfrag{x}[]{$M_{t\bar t}$ [GeV]}
\psfrag{d}[][][0.85]{\qquad $ \;\;\;\;A^{t \bar{t}}_{\text{FB}}(M_{t\bar t})$}
\psfrag{e}[][][0.85]{\qquad\;\; \;$(d \sigma/dM_{t \bar{t}})_R$}
\includegraphics[width=0.5\textwidth]{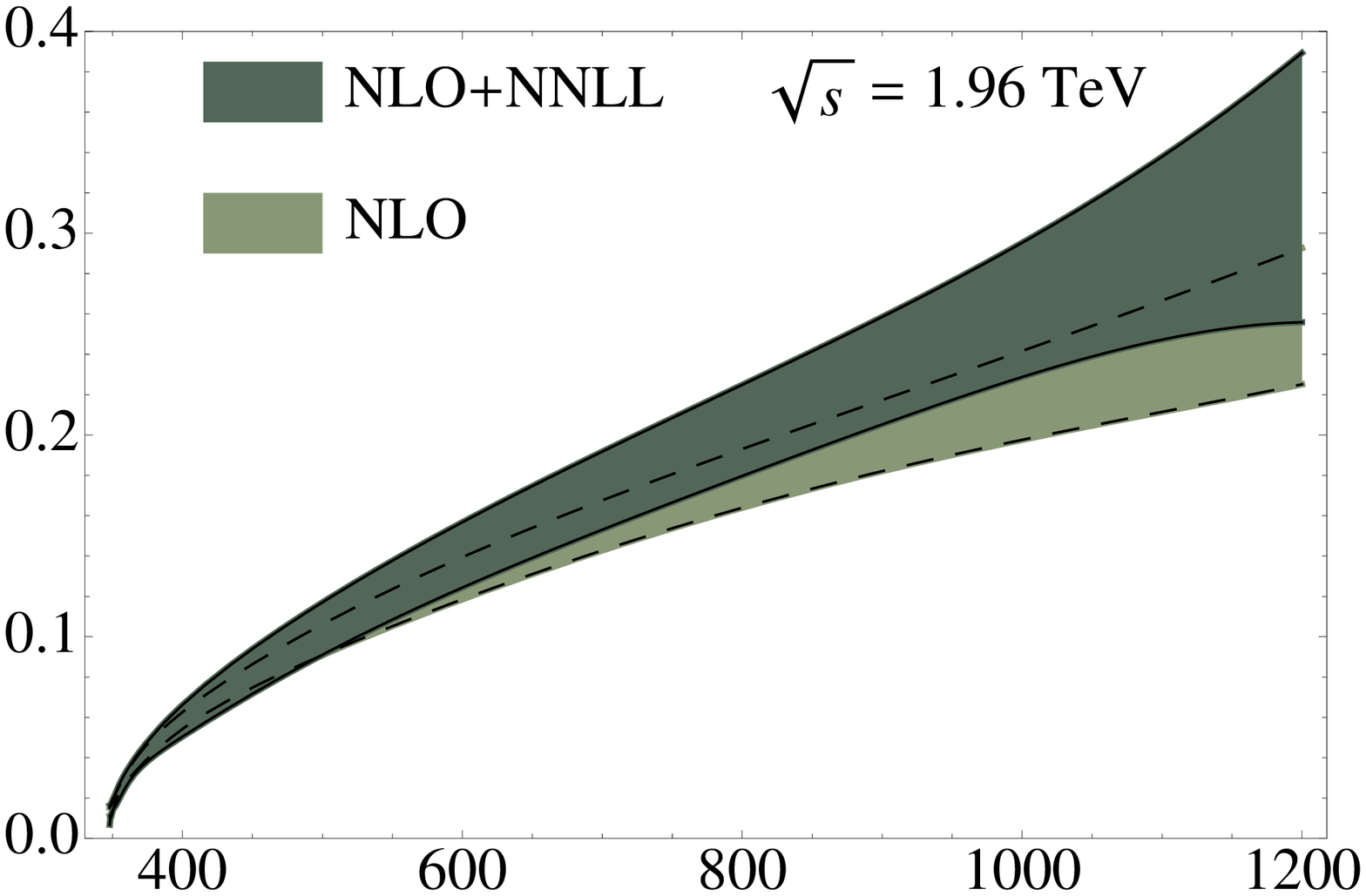}
\end{tabular}
\end{center}
\vspace{-2mm}
\caption{\label{fig:asymplot}Left: The asymmetric cross section 
$d\Delta\sigma^{t\bar{t}}_{\text{FB}}/dM_{t\bar{t}}$ as
a function of the invariant mass at NLO and NLO+NNLL order. Right: 
The asymmetry $A^{t \bar{t}}_{\text{FB}}(M_{t\bar{t}})$.
The bands show the uncertainties related to scale variation as explained in the text.}
\end{figure}

As mentioned in the introduction, the recent measurement of the asymmetry at high values of the
pair invariant mass shows a large deviation from the value predicted by QCD at 
NLO. In order to examine the effects of soft-gluon resummation
on this observable, we first extract from (\ref{eq:fbasymtt}) the invariant-mass 
dependent asymmetry as 
\begin{align}
  \label{eq:fbadiff}
   A^{t\bar{t}}_{\text{FB}}(M_{t\bar{t}}) = \frac{\displaystyle \left(
      \frac{d\sigma}{dM_{t\bar{t}}} \right)_F - \left( \frac{d\sigma}{dM_{t\bar{t}}}
    \right)_B}{\displaystyle \left( \frac{d\sigma}{dM_{t\bar{t}}} \right)_F + \left(
      \frac{d\sigma}{dM_{t\bar{t}}} \right)_B } \,= \, \frac{\displaystyle\frac{d\Delta\sigma^{t\bar{t}}_{\text{FB}}}{dM_{t\bar{t}}}}{
      \displaystyle\frac{d\sigma}{dM_{t\bar{t}}}}\; .
\end{align}
Results for  $A^{t\bar{t}}_{\text{FB}}(M_{t\bar{t}})$ and
$d\Delta\sigma^{t\bar{t}}_{\text{FB}}/dM_{t\bar{t}}$ at NLO and NLO+NNLL order 
are shown in Figure~\ref{fig:asymplot}, where the 
bands reflect uncertainties originating  from scale variations.
The figure shows that the asymmetry increases with the invariant mass and can reach
nearly 40\% at $M_{t\bar{t}}=1200$~GeV. These results are obtained with the default MSTW2008
PDFs and  do not include PDF uncertainties. An analysis shows that the relative PDF error for 
$d\Delta\sigma^{t\bar{t}}_{\text{FB}}/dM_{t\bar{t}}$ increases slightly with
increasing $M_{t\bar{t}}$, from  $7$\% ($400$~GeV) to $9$\% ($1200$~GeV). In
contrast,  the relative PDF error for $A^{t\bar{t}}_{\text{FB}}(M_{t\bar{t}})$ 
is rather small and even decreases with $M_{t\bar{t}}$,  from around $2$\%
($400$~GeV) to $1$\% ($1200$~GeV). 

It is a well-known fact that electroweak corrections start to become
more important for the differential cross section at high pair
invariant mass, due to the presence of Sudakov logarithms. For
instance, at $M_{t\bar t}\sim 1$~TeV, the electroweak corrections to the
differential distribution at the Tevatron are roughly at the -5\%
level \cite{Bernreuther:2006vg, Kuhn:2006vh}, even though for the
total cross section they are negligible.  On the other hand, the electroweak
corrections to the asymmetric cross section given in
\cite{Kuhn:1998kw} do not contain Sudakov logarithms, and an estimate
shows that their distortion of the QCD contribution is roughly independent 
of the invariant mass. We therefore do not expect the
electroweak corrections to significantly alter our results for the 
FB asymmetry even at high values of pair invariant mass.

%
%
%
%
\begin{figure}[h]
\begin{center}
\psfrag{y}[][][1][90]{$A^{t \bar{t}}_{\text{FB}}$}
\psfrag{z}[][][1][90]{$A_{t\bar t}$ [\%]}
\psfrag{x}[]{$M_{t\bar t}$ [GeV]}
\psfrag{d}[][][0.85]{$\sqrt{s}=1.96$\,TeV}
\includegraphics[width=0.5\textwidth]{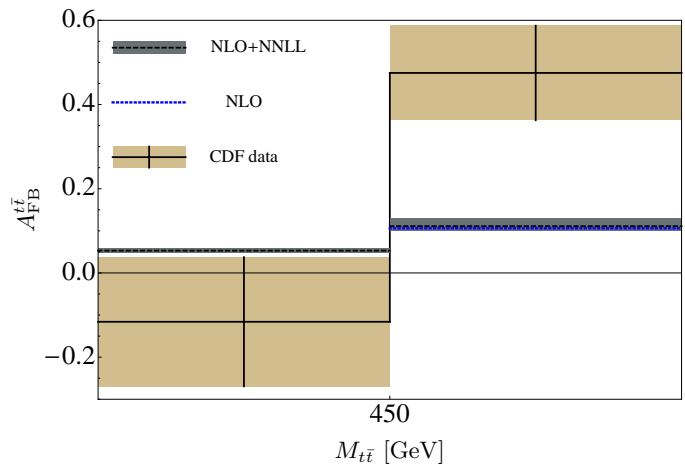}
\end{center}
\vspace{-2mm}
\caption{\label{fig:cdfcompare} The asymmetry in the high and low invariant-mass region as measured
  in \cite{Aaltonen:2011kc},  compared to our predictions at  NLO+NNLL order.
The bands in the NLO+NNLL results are related to uncertainties from 
scale variation, while the NLO result in the higher bin is 
evaluated at $\mu_f= m_t$. }
\end{figure}
%

We hope that better statistics can eventually lead 
to a detailed comparison of experimental results with the asymmetry curve in 
Figure~\ref{fig:asymplot}. At present this is not possible, but 
the CDF collaboration has measured the invariant-mass dependent asymmetry
by separating the events into a high invariant-mass bin ($M_{t\bar{t}}
\ge 450$~GeV)  and a low invariant-mass bin ($M_{t\bar{t}} \le 450$~GeV) \cite{Aaltonen:2011kc}. 
From Figure~\ref{fig:asymplot}, one can see that this choice roughly divides
the total asymmetric cross section equally between the two bins, although 
most of the asymmetric cross section in the high invariant-mass bin originates
from the region close to $450$~GeV.  More precisely, the region 
$450-600\,(700)$~GeV captures more than $75\,(90)$\% of the total asymmetric
cross section in the high invariant-mass bin. 
To compare with the CDF results, we evaluate the binned asymmetry
\begin{align}
  \label{eq:fbabin}
   A^{t\bar{t}}_{\text{FB}}(m_1, m_2) = \frac{\displaystyle 
\int_{m_1}^{m_2} dM_{t\bar{t}}  \left(d\Delta\sigma^{t\bar{t}}_{\text{FB}}/dM_{t\bar{t}}
   \right)}{\displaystyle \int_{m_1}^{m_2} dM_{t\bar{t}}   \left(d\sigma/dM_{t\bar{t}}
    \right) } \; ,
\end{align}
for $M_{t\bar{t}} \le 450$~GeV and for $M_{t\bar{t}} \ge 450$~GeV.  Our
findings are given in Table~\ref{tab:cdfcompare}, along with 
their visual representation in
Figure~\ref{fig:cdfcompare}, which shows the NLO+NNLL
calculation with an error band from scale variations along with the default NLO number in the high invariant-mass
bin. In both bins,
the NLO+NNLL predictions for the asymmetric cross sections 
have considerably smaller scale uncertainties than
the NLO ones, but the results for the 
FB asymmetries are essentially unchanged. As with all other 
results obtained in the $t\bar t$ frame, the scale uncertainties in the 
FB asymmetries are larger in the NLO+NNLL calculation that at NLO.
However, if we had not expanded the ratio, the predicted FB asymmetry in the high invariant-mass bin would be 
9.0\% at NLO and 10.6\% at NLO+NNLL order\footnote{Using MSTW2008 PDFs as an example.}, showing the stability 
of the resummed results under this change of systematics.  

\begin{table}[h]
  \centering
  \begin{tabular}{|l|c|c|c|c|c|c|c|c|c|c|c|c|}
    \hline
    & \multicolumn{6}{c|}{$M_{t\bar{t}} \le 450$~GeV} & \multicolumn{6}{c|}{$M_{t\bar{t}} > 450$~GeV}
    \\ \cline{2-13}
    & \multicolumn{3}{c|}{$\Delta\sigma^{t\bar{t}}_{\text{FB}}$ [pb]} & \multicolumn{3}{c|}{$A^{t\bar{t}}_{\text{FB}}$[\%]}
    & \multicolumn{3}{c|}{$\Delta\sigma^{t \bar{t}}_{\text{FB}}$ [pb]} & \multicolumn{3}{c|}{$A^{t \bar{t}}_{\text{FB}}$[\%]}
    \\ \hline
    CDF
    & \multicolumn{3}{c|}{}
    & \multicolumn{3}{c|}{$-11.6${\footnotesize $^{+15.3}_{-15.3}$}}
    & \multicolumn{3}{c|}{}
    & \multicolumn{3}{c|}{47.5{\footnotesize $^{+11.2}_{-11.2}$}}
    \\ \hline 
    & MSTW & CTEQ & NNPDF    & MSTW & CTEQ & NNPDF    & MSTW & CTEQ & NNPDF    & MSTW & CTEQ & NNPDF
    \\ \hline
    NLO
    & 0.17{\footnotesize $^{+0.10}_{-0.05}$} & 0.18{\footnotesize $^{+0.09}_{-0.05}$} & 0.19{\footnotesize $^{+0.09}_{-0.06}$}
    & 5.2{\footnotesize $^{+0.6}_{-0.2}$} & 5.3{\footnotesize $^{+0.4}_{-0.4}$} & 5.4{\footnotesize $^{+0.3}_{-0.4}$}
    & 0.22{\footnotesize $^{+0.13}_{-0.07}$} & 0.22{\footnotesize $^{+0.12}_{-0.07}$} & 0.23{\footnotesize $^{+0.12}_{-0.07}$}
    & 10.8{\footnotesize $^{+1.0}_{-0.8}$} & 10.4{\footnotesize $^{+1.0}_{-0.6}$} & 10.9{\footnotesize $^{+0.7}_{-0.6}$}
    \\ \hline 
    NLO+NNLL
    & 0.21{\footnotesize $^{+0.04}_{-0.03}$} & 0.22{\footnotesize $^{+0.04}_{-0.04}$} & 0.23{\footnotesize $^{+0.04}_{-0.04}$}
    & 5.2{\footnotesize $^{+0.9}_{-0.6}$} & 5.2{\footnotesize $^{+0.8}_{-0.6}$} & 5.4{\footnotesize $^{+0.7}_{-0.6}$}
    & 0.24{\footnotesize $^{+0.04}_{-0.04}$} & 0.25{\footnotesize $^{+0.05}_{-0.04}$} & 0.26{\footnotesize $^{+0.04}_{-0.04}$}
    & 11.1{\footnotesize $^{+1.7}_{-0.9}$} & 10.8{\footnotesize $^{+1.7}_{-0.9}$} & 11.4{\footnotesize $^{+1.3}_{-1.0}$}
    \\ \hline  
  \end{tabular}
  \caption{\label{tab:cdfcompare} Comparison of the low- and high-mass asymmetry
    $A^{t\bar{t}}_{\text{FB}}$ with CDF data \cite{Aaltonen:2011kc}, along with results
    for the asymmetric cross section. The errors in the QCD predictions refer to
    perturbative  uncertainties related to scale variation.}
\end{table}
%
%
%
\begin{figure}
\begin{center}
\begin{tabular}{ll}
\psfrag{y}[][][1][90]{$d\Delta\sigma^{t\bar{t}}_{\text{FB}}/d\Delta y$\;[pb]}
\psfrag{x}[]{$\Delta y$}
\psfrag{d}[][][0.90]{$\sqrt{s}=1.96$\,TeV}
\includegraphics[width=0.5\textwidth]{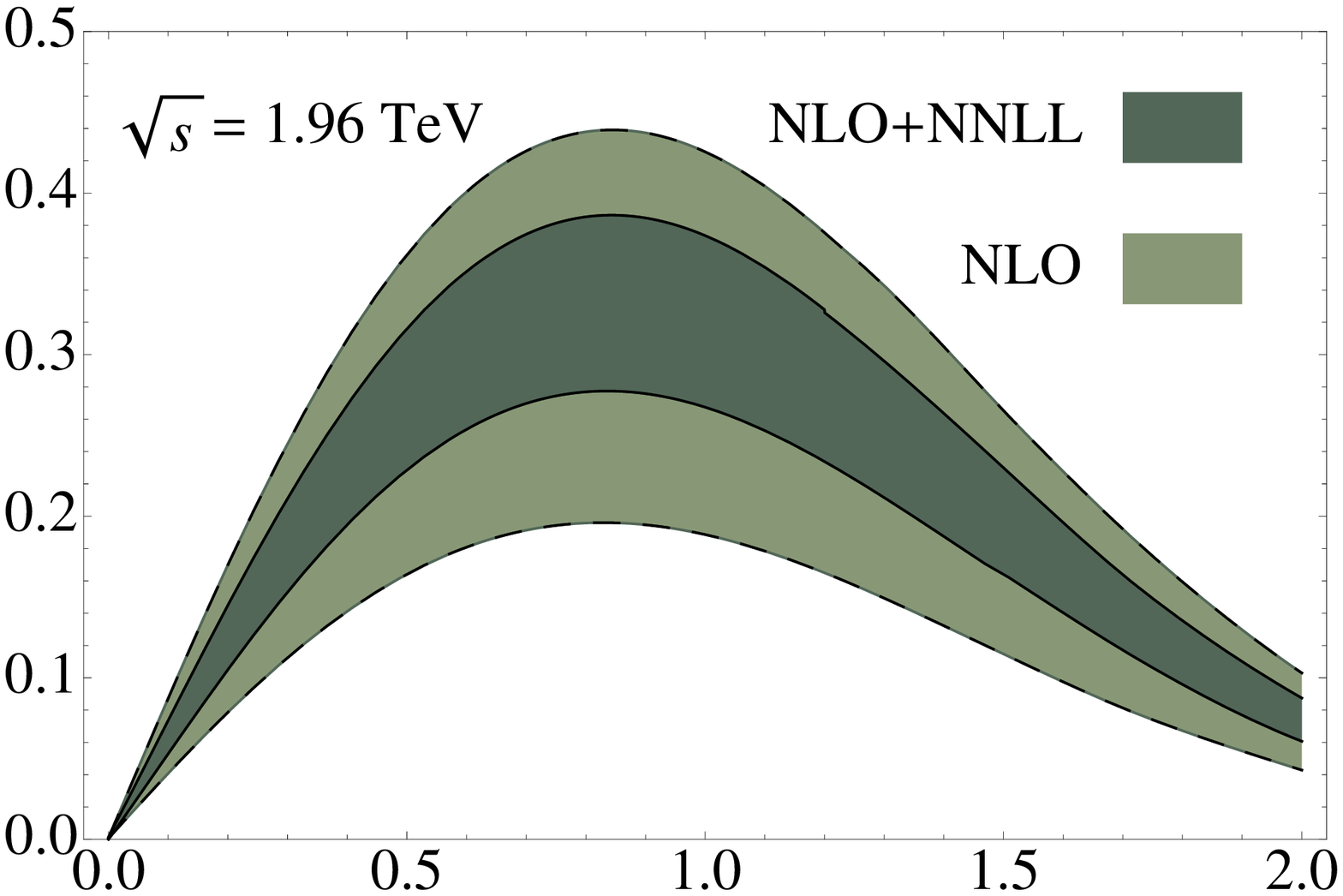}
\psfrag{y}[][][1][90]{$A^{t \bar{t}}_{\text{FB}}(\Delta y)$}
\psfrag{x}[]{$\Delta y$}
\psfrag{d}[][][0.90]{$\sqrt{s}=1.96$\,TeV}
\includegraphics[width=0.5\textwidth]{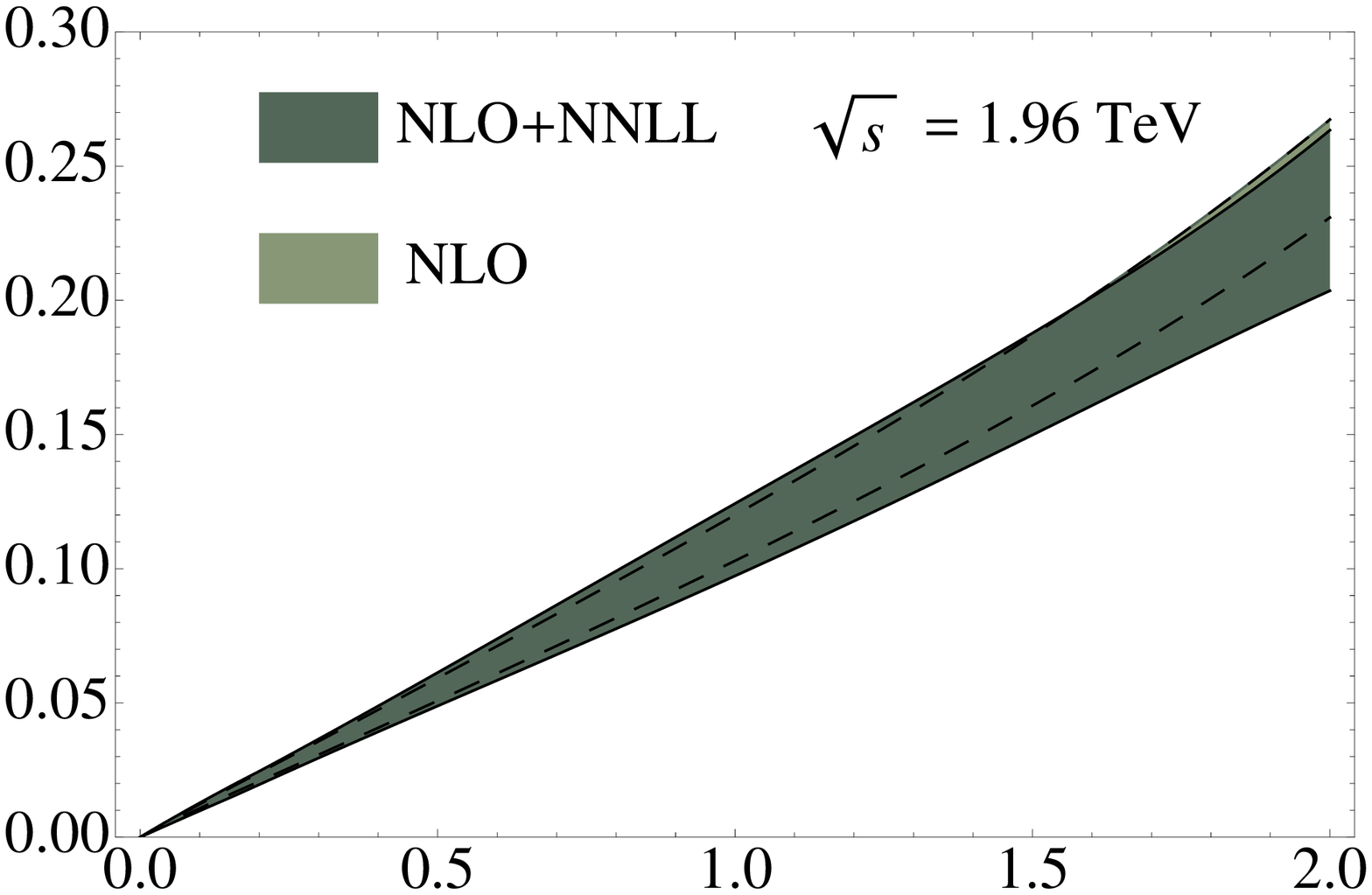}
\end{tabular}
\end{center}
\vspace{-2mm}
\caption{\label{fig:rapidity2} Left: The asymmetric differential cross section $ d\Delta\sigma^{t
      \bar{t}}_{\text{FB}}/d\Delta y$. Right: The asymmetry $A^{t \bar{t}}_{\text{FB}}(\Delta y)$.
  The bands show the errors related to scale variation as explained in the text.}
\end{figure}

We now turn to a discussion of the PDF uncertainties in the binned results, in this case
deviating slightly from our usual procedure. The reason is that to compute the PDF
uncertainties for the binned asymmetry at NLO+NNLL order, we need to run the Monte Carlo
program MadGraph at NLO for each of the PDFs in the error set, which is rather time
consuming. As a compromise, we have estimated the PDF uncertainties using only the pieces
of the NLO calculation which are leading in the threshold limit $M_{t\bar t}^2/\hat{s}\to
1$, as obtained in \cite{Ahrens:2010zv}. Since these leading pieces alone account for the
bulk of the NLO FB asymmetry, the relative PDF uncertainties obtained from these terms
should provide a good approximation to those in the full NLO and NLO+NNLL results. For the
MSTW2008 set, we find a relative PDF uncertainty of about $7\%$ for the asymmetric cross
section and about $2$\% for the FB asymmetry at 90\% CL, in both the low and high
invariant-mass bins.

Our calculations show that neither higher-order corrections from soft-gluon
resummation nor the inclusion of a systematic uncertainty coming from PDF
usage reduces in any significant way the current discrepancy between theory 
and experiment for the FB asymmetry in the high invariant-mass bin, which remains 
above the $3\sigma$ level when using our NLO+NNLL calculations.

\subsection{Rapidity-dependent asymmetry}

A further observable of interest is the rapidity dependence
of the FB asymmetry in the $t\bar t$ frame. 
In practice, experiments measure the
asymmetry as a function of the pair rapidity difference 
$\Delta y = y_t - y_{\bar t}$ \cite{Aaltonen:2011kc}. 
We can calculate the differential cross 
section in this variable from the results in PIM kinematics by 
using that, up to power corrections which vanish in the soft limit, 
\begin{align}
\Delta y = \ln\left(\frac{1+  \cos\theta\sqrt{1-4m_t^2/M_{t\bar t}^2}}{1-  \cos\theta\sqrt{1-4m_t^2/M_{t\bar t}^2}}\right)  .
\end{align} 
After changing variables from the scattering angle to the pair 
rapidity difference, we express the asymmetric cross section as 
\begin{align}
  \label{eq:fbasymttrap}
  \Delta\sigma^{t\bar{t}}_{\text{FB}} &= \int_0^{\Delta y_+} d\Delta y \left[
    \int_{M_{t\bar t}^{\rm min}}^{\sqrt{s}} dM_{t\bar t} \, 
\frac{d^2 \sigma^{p\bar{p} \to t\bar t X
     }}{dM_{t\bar t}d\Delta y} -\int_{M_{t\bar t}^{\rm min}}^{\sqrt{s}} dM_{t\bar t} 
\,\frac{d^2 \sigma^{p\bar{p} \to t\bar t X
     }}{dM_{t\bar t}d\Delta \bar y} \Bigg|_{\Delta \bar{y}=-\Delta y} \right]
  \nonumber
  \\
  &\equiv \int_0^{\Delta y_+} d\Delta y \left[ \left( \frac{d\sigma}{d\Delta y} \right)_F - \left(
      \frac{d\sigma}{d\Delta y} \right)_B \right]\equiv \int_0^{\Delta y_+} d\Delta y  \frac{d \Delta\sigma^{t\bar{t}}_{\text{FB}}}{d\Delta y} \,,
\end{align}
where
\begin{align}
 \Delta y_+ =  \ln\left(\frac{1+\sqrt{1-4m_t^2/s}}{1-\sqrt{1-4m_t^2/s}}\right) \quad \text{and}
  \quad M_{t\bar t}^{\text{min}} = 2m_t \cosh \left(\Delta y/2\right)
  \,.
\end{align}
Using these definitions, we can also introduce the  
$\Delta y$-dependent asymmetry
\begin{align}
  \label{eq:fbaYttdiff}
  A^{t\bar t}_{\text{FB}}(\Delta y) 
  = \frac{\displaystyle  \frac{d \Delta \sigma^{t\bar{t}}_{\text{FB}}}{d\Delta y}
   }{\displaystyle
    \left(\frac{d\sigma}{d\Delta y}\right)_F + \left(\frac{d\sigma}{d\Delta y}\right)_B} \, ,
\end{align}
and binned asymmetries analogous to  (\ref{eq:fbabin}), where 
the numerator and denominator of the above
expression are integrated over a range in $\Delta y$. 
Note that the integration region above implies that higher values of 
$\Delta y$  correspond to higher values of $M_{t\bar t}$.  For example, the 
restriction $\Delta y>1$ used in the binned analysis below corresponds to 
events with $M_{t\bar t}>390$~GeV.  

%
\begin{figure}[t]
\begin{center}
\psfrag{y}[][][1][90]{$A^{t \bar{t}}_{\text{FB}}$}
\psfrag{z}[][][1][90]{$A_{t\bar t}$ [\%]}
\psfrag{x}[]{$\Delta y$}
\psfrag{d}[][][0.85]{$\sqrt{s}=1.96$\,TeV}
\includegraphics[width=0.5\textwidth]{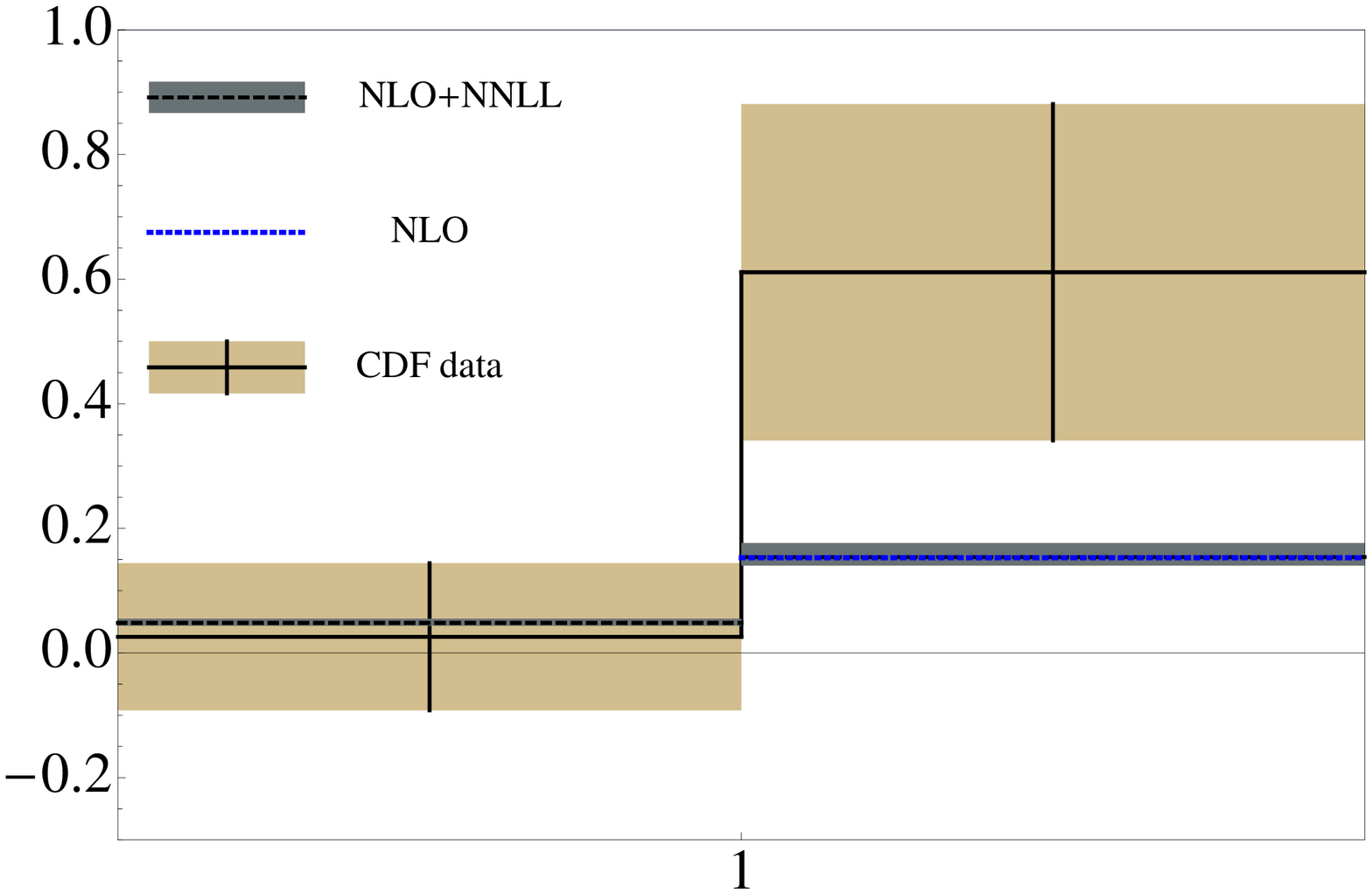}
\end{center}
\vspace{-2mm}
\caption{\label{fig:cdfcompare2} The asymmetry for $\Delta y < 1$ and $\Delta y \ge 1$ as measured
  in \cite{Aaltonen:2011kc}, compared to our predictions at NLO+NNLL order. 
The bands in the NLO+NNLL results are related to uncertainties from 
scale variation, while the NLO result in the higher bin is 
evaluated at $\mu_f= m_t$. }
\end{figure}
%
%

\begin{table}[t]
  \centering
  \begin{tabular}{|l|c|c|c|c|}
    \hline
    & \multicolumn{2}{c|}{$\Delta y < 1 $} & \multicolumn{2}{c|}{$\Delta y \ge 1$}
     \\ \cline{2-5}
    & $\Delta\sigma^{t \bar{t}}_{\rm FB}$ [pb] & $A^{t \bar{t}}_{\text{FB}}$ [\%]  & $\Delta\sigma^{t \bar{t}}_{\rm FB}$ [pb]  &  $A^{t \bar{t}}_{\text{FB}}$ [\%]
    \\ \hline
CDF & & $2.6^{+11.8}_{-11.8}$ & & $61.1^{+25.6}_{-25.6}$   
    \\ \hline
    NLO 
    & 0.204{\footnotesize $^{+0.105}_{-0.064}$} 
    & 4.86{\footnotesize $^{+0.42}_{-0.35}$}
    & 0.172{\footnotesize $^{+0.094}_{-0.057}$} 
    & 15.29{\footnotesize $^{+1.26}_{-1.11}$}
     \\ \hline 
    NLO+NNLL
    & 0.230{\footnotesize $^{+0.040}_{-0.035}$} 
    & 4.77{\footnotesize $^{+0.39}_{-0.35}$}
    & 0.196{\footnotesize $^{+0.035}_{-0.031}$} 
    & 14.59{\footnotesize $^{+2.16}_{-1.30}$}
     \\ \hline  
  \end{tabular}

  \caption{\label{tab:ttbar2} Comparison of  $A^{t\bar{t}}_{\text{FB}}$ for
    $\Delta y < 1$ and $\Delta y \ge 1$ with CDF data \cite{Aaltonen:2011kc},
along  with the asymmetric cross section.  The errors in the QCD predictions
refer to the uncertainties related to scale variation.}
\end{table}

We show results related to the rapidity dependence of the FB asymmetry
in Figures~\ref{fig:rapidity2} and \ref{fig:cdfcompare2},  and in 
Table~\ref{tab:ttbar2}. In all cases we use MSTW2008 PDFs.
The more detailed results in the Figure~\ref{fig:rapidity2} show the 
differential asymmetric cross section along 
with the pair-rapidity dependent FB asymmetry, in the form of bands 
related to uncertainties from scale variations.  While resummation stabilizes
the asymmetric cross section compared to NLO, there is little effect on the 
FB asymmetry. The results for the binned asymmetry are given in Table~\ref{tab:ttbar2},
along with their visual representation in Figure~\ref{fig:cdfcompare2}, which 
shows the NLO+NNLL result with an error band from scale variations along with the default NLO number in the 
higher bin.
For events where $\Delta y \le 1$, the QCD prediction is in agreement with the CDF
measurement \cite{Aaltonen:2011kc}. In the bin where $\Delta y \ge 1$, 
the predicted asymmetry is lower than the measured one by $\sim 1.5\sigma$. 
Again in this case, soft-gluon resummation changes the NLO predictions only slightly.

\section{Charge asymmetry at the LHC}

The Tevatron results for the FB asymmetry at high pair invariant mass 
and rapidity  hint at a discrepancy with the Standard Model. 
It would of course be desirable to study the physics responsible
for this effect through measurements at the LHC.  

The total and differential FB asymmetries at the LHC vanish, because of the
symmetric initial state.  However, while charge conjugation invariance of the 
strong interaction implies that the
rate for the forward production of top-quarks is equal to the rate for backward
production of antitop quarks at the Tevatron, this is not the case at the
LHC. At a proton-proton collider the total rate for top and antitop production
in the forward or backward hemisphere is equal, but at a given rapidity the
rates differ. In fact, at large (small) rapidities the rate for top-quark
production is noticeably larger (smaller) than that for antitop production
\cite{Kuhn:1998kw}, so although there is no FB asymmetry at the LHC there is 
a differential charge asymmetry. Like the FB asymmetry at the Tevatron, 
this charge asymmetry at the LHC is related to the asymmetric 
part of the $q\bar q$ partonic cross section, implying a direct 
correlation between potential new physics contributions to the 
two measurements.  The charge asymmetry at the LHC is generally
smaller than the FB asymmetry at the Tevatron due to large contributions
from the $gg$ channel to the charge-symmetric part of the differential cross 
section, but the rapidity reach at the LHC is larger and the charge asymmetry thus
provides complementary information.

\begin{figure}[t]
\begin{center}
\begin{tabular}{ll}
\psfrag{y}[][][1][90]{$\Delta\sigma^{pp}_{\text{C}}(y_{\rm cut})$\;[pb]}
\psfrag{x}[]{$y_{\rm cut}$}
\psfrag{d}[][][0.90]{$\sqrt{s}=7$\,TeV}
\includegraphics[width=0.5\textwidth]{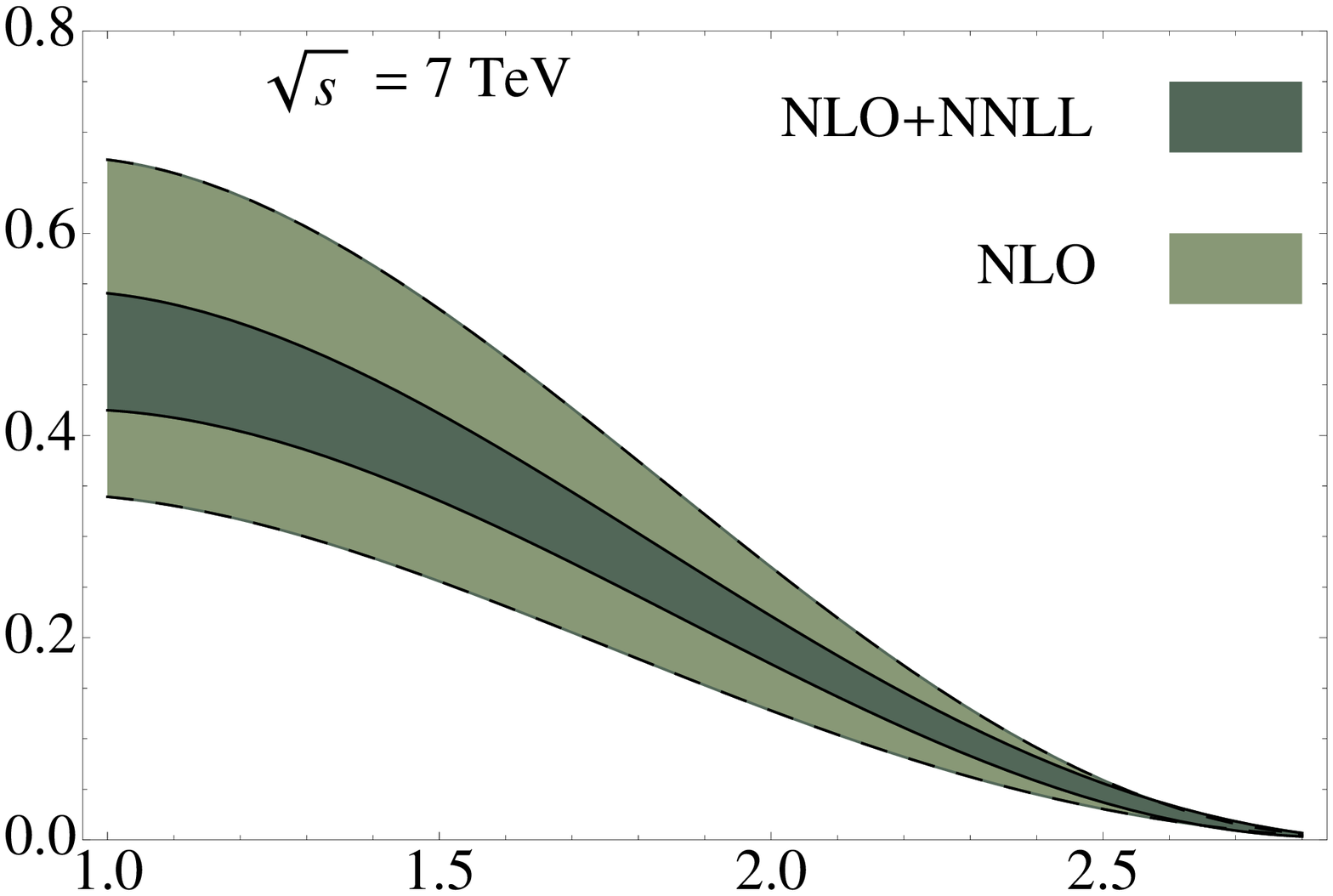}
\psfrag{y}[][][1][90]{$A^{pp}_{\text{C}}(y_{\rm cut})$\;[\%]}
\psfrag{x}[]{$y_{\rm cut}$}
\psfrag{d}[][][0.90]{$\sqrt{s}=7$\,TeV}
\includegraphics[width=0.5\textwidth]{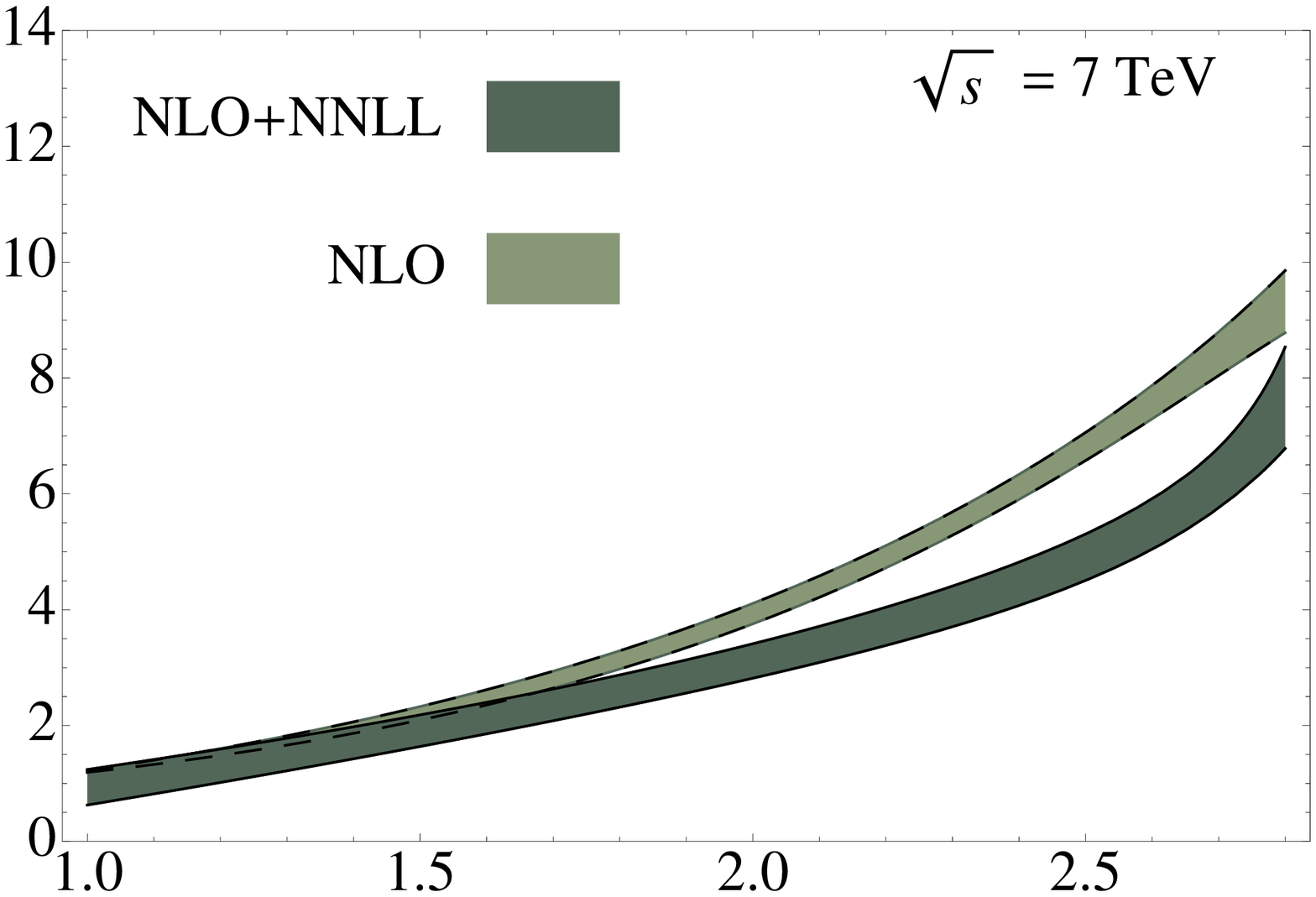}
\end{tabular}
\end{center}
\vspace{-2mm}
\caption{\label{fig:LHCasym} Left: The partially integrated charge-asymmetric
cross section $\Delta \sigma^{pp}_{\text{C}}(y_{\rm cut})$. Right: 
The partially integrated charge asymmetry $A^{p p}_{\text{C}}(y_{\rm cut})$.
The bands show the uncertainties related to scale variation.}
\end{figure}

In this section we study the simplest realization of a charge asymmetry  
at the LHC, namely the rapidity-dependent quantity in the laboratory ($pp$) 
frame. In particular, we focus on the partially integrated charge asymmetry
and charge-asymmetric cross section, where we impose the 
restriction $y>y_{\rm cut}$ on the differential cross section. 
We define these through
\begin{align}
\label{eq:lhcasym}
 A^{p p}_{\text{C}}(y_{\rm cut})
  = \frac{\displaystyle   \int_{y_{\rm cut}}^{y^+_t} dy_t \left( \frac{d \sigma^
{p p \to t
        X_{\bar{t}}}}{dy_t} -
\frac{d \sigma^{pp \to \bar{t} X_{t}}}{dy_{\bar{t}}} \Bigg|_{y_{\bar{t}}=y_t} \right
)
    }{\displaystyle
   \int_{y_{\rm cut}}^{y^+_t} dy_t \left( \frac{d \sigma^{p p \to t
        X_{\bar{t}}}}{dy_t} +
\frac{d \sigma^{pp \to \bar{t} X_{t}}}{dy_{\bar{t}}} \Bigg|_{y_{\bar{t}}=y_t}
  \right)}
\equiv
\frac{ \Delta \sigma^{pp}_{\rm{C}}(y_{\rm cut})  }{\displaystyle
 \int_{y_{\rm cut}}^{y^+_t} dy_t \left( \frac{d \sigma^{p p \to t
        X_{\bar{t}}}}{dy_t} +
\frac{d \sigma^{pp \to \bar{t} X_{t}}}{dy_{\bar{t}}} \Bigg|_{y_{\bar{t}}=y_t}
  \right)  } \, ,
\end{align}
with $y_t^+$ as in (\ref{eq:yplimits}). 

We can study the partially integrated charge asymmetry and asymmetric cross
section (\ref{eq:lhcasym}) at NLO and NLO+NNLL order using the results for the
differential cross section in 1PI kinematics obtained in \cite{Ahrens:2011mw}.
The results generated with MSTW2008 PDFs are shown in
Figure~\ref{fig:LHCasym}, where the bands reflect uncertainties related to
scale variations carried out with the same procedure as at the Tevatron.  For
the charge-asymmetric cross section shown in the left-hand panel of the figure,
the main effect of the resummation is to decrease the scale dependence of the
result to a relatively small region of the NLO error band. For the partially
integrated asymmetry shown in the right-hand panel of the figure, the
resummation is a mild effect up to $y_{\rm cut}\sim 1.5$, but substantially
reduces the asymmetry at higher values of the cut. However, due to large K
factors in the gluon channel at the LHC, the uncertainty band for the NLO
curve is very sensitive to whether one consistently expands the asymmetry
ratio in (\ref{eq:lhcasym}). If we had deviated from our normal procedure and
had not expanded the ratio, instead evaluating the denominator at NLO order, the NLO
band would actually overlap quite well with the NLO+NNLL results shown in
figure.  The NLO+NNLL result is largely insensitive to this change of
systematics--the result where both the numerator and denominator are evaluated
at NLO+NNLL order is within the error band shown in the figure.

The partially integrated charge asymmetry vanishes for $y_{\rm cut}=0$, and becomes
progressively larger at higher values of the cut.  However, the charge
asymmetric cross section shown in the left-hand side of the figure is very
small at higher rapidity values and the experimental measurement is
difficult.  A reasonable way to compare theory and experiment in this case
would be to perform a measurement in a high-rapidity bin with $y>y_{\rm
cut}\sim 1\mbox{-}1.5$.  In such
a bin the Standard Model charge asymmetry is predicted to be only slightly
different from zero, less than 2\% depending on the exact choice of the cut,
so any appreciable charge asymmetry in an experimental measurement would be a
clear sign of a new physics contribution to the high-rapidity region of the
distribution, which is already hinted at by the Tevatron measurements.

Note that other partially integrated differential distributions
can be used to study the phase-space dependence of the charge asymmetry.
For instance, one can consider the so-called central charge asymmetry 
introduced in  \cite{Antunano:2007da}, which imposes a cut 
$|y| <y_{\rm C}$ along with a restriction to relatively high pair invariant
mass in order  to reduce the contribution from the  
$gg$ channel to the symmetric part of the partially integrated cross section. 
Unfortunately, it is beyond the scope of the paper to perform 
an NLO+NNLL analysis for such an observable, because it requires a combination
of PIM and 1PI kinematics which is not possible to derive starting
directly from the results of \cite{Ahrens:2010zv, Ahrens:2011mw}.

\section{Conclusions}

The total top-quark FB asymmetry measured at the Tevatron is not in good
agreement with the predictions obtained at the first non-vanishing order in perturbative
QCD. The tension between theory and experiment is about two standard
deviations for the asymmetry measurement in the laboratory frame, and
approximately 1$\sigma$ in the top-pair rest frame.  Moreover,
the measurement of the asymmetry in the large pair invariant-mass region
($M_{t\bar{t}} \ge 450$~GeV) by the CDF collaboration \cite{Aaltonen:2011kc}
differs from the leading-order prediction in QCD by more that three standard
deviations, and there is also a discrepancy between theory and experiment for the
FB asymmetry at values of the top-pair rapidity difference $\Delta y >1$.  

The calculations of the double differential distribution for the top-quark pair
production in two different kinematic schemes, carried out in
\cite{Ahrens:2010zv, Ahrens:2011mw}, allow us to evaluate the FB 
asymmetry in both the laboratory frame and in the top-pair rest frame.
The predictions obtained in this way include the full NLO corrections 
to the total and asymmetric cross sections, whose ratio determines the 
FB asymmetry, as well as the resummation of soft-gluon emission effects up to
NNLL accuracy. Such NLO+NNLL results represent the most accurate determination 
of the QCD contribution to the asymmetry that can be obtained to date.
Studies of the NLO results indicated that power corrections to our results
obtained in the soft limit within the effective theory approach are small,
due to the mechanism of dynamical threshold enhancement.  
The full NNLO calculation would be very useful in quantifying this
at the next order in perturbation theory, but we do not expect such
power corrections to change our predictions in any significant way.

Tables~\ref{tab:ppbar} and \ref{tab:ttbar} show that the numerical impact of
the NNLL corrections on the total asymmetry is  modest,  less than $3 \%$ of
the central value in all cases considered. 
It must be observed that while the total and asymmetric cross sections
calculated at NLO+NNLL order have smaller scale uncertainties than the
corresponding quantities calculated at NLO, the same is not always true for the FB
asymmetry, indicating that the errors at NLO are affected by accidental
cancellations and do not reflect the true perturbative uncertainties.  
After the inclusion of the NNLL corrections, the predicted values of the 
asymmetry remain lower than the measured ones by $\sim 2 \sigma$ in 
the $p \bar{p}$ frame and $\sim 1 \sigma$ in the $t \bar{t}$ frame.
We have also shown that PDF uncertainties are much reduced for the FB
asymmetries compared to those for the asymmetric or total cross sections.  For instance, 
the total asymmetry calculated with MSTW2008 PDFs differs from the one
obtained from CTEQ6.6 or NNPDF2.1 PDFs by only a few percent.

The NLO+NNLL calculation of the asymmetry in the high invariant-mass bin $M_{t
\bar{t}} \ge 450$~GeV increases the NLO prediction only by a very small
amount, about 3\% of the central value.  The perturbative and PDF
uncertainties of the calculation are much smaller than the experimental
errors.  Hence, if the Tevatron measurement of the FB asymmetry in the high
invariant-mass bin is confirmed as the experimental error is
reduced, it will become a compelling signal for physics beyond the Standard
Model.

Given the situation at the Tevatron, it is well motivated to
measure observables sensitive to the top quark asymmetry also at the LHC. 
Since the FB asymmetry vanishes at the
LHC, one of the possible choices consists in measuring the charge asymmetry
partially integrated in a given rapidity region. Starting from the NLO+NNLL
resummed rapidity distributions for the top and antitop quarks, we calculated
the charge asymmetry as a function of a lower rapidity cut in the laboratory frame.  A careful choice
of the lower bound on the rapidity integration region can allow one to use
this observable to test for the presence of new physics in the top-quark sector at
the LHC, thus providing complementary information to the Tevatron measurements.

\vspace{2mm}

{\em Acknowledgments:\/} We would like to thank Rikkert Frederix for providing us a
private version of MadFKS. This research was supported in part by the State of
Rhineland-Palatinate via the Research Center {\em Elementary Forces and Mathematical
  Foundations}, by the German Federal Ministry for Education and Research under grant
05H09UME, by the German Research Foundation under grant NE398/3-1, by the European
Commission through the {\em LHCPhenoNet\/} Initial Training Network PITN-GA-2010-264564,
and by the Schweizer Nationalfonds under grant 200020-124773.

\end{document}